\documentclass[preprint,showpacs,preprintnumbers,amsmath,amssymb]{revtex4}
\usepackage[utf8]{inputenc}
\usepackage{amsmath}
\usepackage{amssymb}
\usepackage[dvips]{graphicx}
\usepackage{dcolumn}
\usepackage{bm}
\usepackage{color}

\begin{document}

\bibliographystyle{apsrev}

\title{Proving the Perdew-Burke-Ernzerhof density functional designed
for metallic bulk and surface systems}

\author{M. Ropo, K. Kokko}
\affiliation{Department of Physics, University of Turku, FIN-20014
Turku, Finland}

\author{L. Vitos}
\affiliation{Applied Materials Physics, Department of
Materials Science and Engineering, Royal Institute of
Technology, SE-10044 Stockholm,Sweden}
\affiliation{Condensed Matter Theory Group, Physics
Department, Uppsala University, SE-75121 Uppsala, Sweden}
\affiliation{Research Institute for Solid State Physics
and Optics,P.O. Box 49, H-1525 Budapest, Hungary}

\date{06 March 2008}

\begin{abstract}
We test the accuracy of the revised Perdew-Burke-Ernzerhof
exchange-correlation density functional (PBEsol) for metallic bulk
and surface systems. It is shown that, on average, PBEsol yields
equilibrium volumes and bulk moduli in close agreement with the
former generalized gradient approximation (PBE) and two gradient
level functionals derived from model system approach (LAG and AM05).
On the other hand, for close-packed metal surfaces, PBEsol has the
same performance as AM05, giving significantly larger surface
energies than PBE and LAG.
\end{abstract}

\maketitle

Today, density functional theory \cite{Hohenberg1964} has become a
state-of-the-art approach in the \emph{ab initio} description of
condensed matter. Its success, to a large extent, may be attributed
to the unanticipated high performance of the local density
approximation (LDA) defined as the zeroth order term of the density
gradient expansion \cite{Kohn1965}. Attempts to go beyond LDA have
led to the elaboration of the gradient corrected functionals. The
pioneering work by Langreth and Mehl \cite{Langreth1981} was
followed by a large number of different approximations \cite{Perdew1986,%
Perdew1986a,Perdew1991,Perdew1996,Zhang1998,Perdew1998,Lee1988,Becke1988,
Hamprecht1998,Hammer1999,Vitos2000,Armiento2005,Perdew2007}. For
computational solid state physics, the first real breakthrough was
the stabilization of the diverging term from the second order
gradient expansion within the so called generalized gradient
approximation (GGA) \cite{Perdew1986,Perdew1986a}. With this early
approach one could recover, e.g., the correct ground state of Fe at
ambient condition. Later incarnations
\cite{Perdew1991,Perdew1996,Hammer1999} refined the GGA with the
main goal of designing a universal functional for atoms and
molecules as well as bulk and surface systems. During the last
decades, among these GGA functionals, the most successful version
has been the PBE functional proposed by Perdew, Burke and Ernzerhof
\cite{Perdew1996}.

An alternative approach for incorporating effects due to
inhomogeneous electron density was put forward by Kohn and Mattsson
\cite{Kohn1998}. In particular, they presented a description for the
electronic edge within the linear potential or Airy gas
approximation. The proposed model was first elaborated by Vitos et
al. \cite{Vitos2000,Vitos2000a} and later further developed by
Armiento and Mattsson \cite{Armiento2005} within the subsystem
functional (SSF) approach \cite{Armiento2002}. Functionals from this
family, by construction, include important surface effects and
therefore are expected to perform well for systems with electronic
surface. In addition, these functionals turned out to be superior,
on average, compared to the common GGA approaches also in bulk
systems \cite{Vitos2000,Armiento2005,Vitos2007,Mattsson2008}.

Most recently, Perdew and co-workers \cite{Perdew2007} have
introduced a new gradient level functional by revising the PBE
functional \cite{Perdew1996} for solids and their surfaces. Keeping
the exact mathematical constrains of PBE, the authors lifted the
orthodox bias toward the atomic energies by restoring the
first-principles gradient expansion for exchange and readjusting the
correlation term using the jellium surface exchange-correlation
energies obtained at meta-GGA level \cite{Tao2003}. We note that
such readjustment of the (LDA level) correlation energy has
originally been proposed by Armiento and Mattsson
\cite{Armiento2005} in their SSF approach. The new GGA functional,
referred to as PBEsol \cite{Perdew2007}, has been designed to yield
improved equilibrium properties of densely-packed solids and, most
importantly, to remedy the deficiencies of the GGA functionals for
surfaces \cite{Perdew1992a,Vitos1998,Liu2006}.

The aim of this work is to establish the accuracy of the PBEsol
exchange-correlation functional in the case of bulk metals and
transition metal surfaces. We have selected $10$ simple metals and
$19$ transition metals for testing the equation of state, and the
4$d$ transition series plus Rb and Sr for testing the surface
energy. For all metals the experimental low-temperature
crystallographic phase has been considered \cite{Young1991}. In
these tests, we compare the performance of the PBEsol functional to
those obtained in LDA, PBE, LAG and AM05 approximations. For LDA, we
use the Perdew and Wang parametrization \cite{Perdew1992} of the
quantum Monte-Carlo data by Ceperley and Alder \cite{Ceperley1980}.
The LAG functional \cite{Vitos2000} is based on the exchange energy
obtained within the Airy gas approximation \cite{Kohn1998} and the
LDA correlation energy \cite{Perdew1992}. The AM05 approximation,
proposed by Armiento and Mattsson \cite{Armiento2005}, goes beyond
the LAG approach by taking into account non-LDA correlation effects
using the jellium surface model. Hence, the main difference between
LAG and AM05, both of them belonging to the SSF class of
functionals, is the surface-like correlation term included in the
latter functional. Extensive tests on the LAG and AM05
approximations for bulk metals can be found in Refs.
\cite{Vitos2000,Armiento2005,Vitos2007,Mattsson2008}.

The present calculations were performed using the exact muffin-tin
orbitals (EMTO) method
\cite{Vitos2007,Vitos2000b,Vitos2001,Andersen1994}. The EMTO method
is a screened Korringa-Kohn-Rostoker method that uses optimized
overlapping muffin-tin potential spheres to represent the
one-electron potential. The total energy was computed at the full
charge density level \cite{Kollar2000}, which has proved to have the
accuracy of full potential techniques \cite{Asato1999}. The
Kohn-Sham equations were solved within the scalar-relativistic and
soft-core approximations. The Green's function was calculated for
$16$ complex energy points distributed exponentially on a
semi-circular contour including the valence states. We employed the
double Taylor expansion approach \cite{Kissavos2007} to get accurate
slope matrix for each energy point. The EMTO basis set included
$s,p,d$ and $f$ states. In bulk calculations, we used $280, 240$ and
$320$ inequivalent $\vec k$-points in the irreducible wedge of the
body centerec cubic (bcc), face centered cubic (fcc) and hexagonal
close-packed (hcp) Brillouin zones, respectively. The equilibrium
volumes and bulk moduli were extracted from the equation of state
(EOS) described by a Morse function \cite{Moruzzi1988} fitted to the
total energies calculated for five different volumes around the
equilibrium.

All self-consistent calculations were carried out within LDA, and
the gradient terms were included in the total energy within the
perturbative approach \cite{Asato1999}. To assess the accuracy of
this approach, we carried out additional fully self-consistent PBE
calculations for bulk bcc Fe and W, and for fcc Cu and Rh. We find
that the average error introduced by the perturbative treatment of
the PBE gradient correction is $\sim 0.07$ Bohr$\times 10^{-2}$ in
the equilibrium atomic radius and $\sim 4$ GPa in the bulk modulus.
These errors are below the numerical accuracy of our calculations.

It has been shown \cite{Vitos1998} that the surface energy
anisotropy shows negligible dependence on the exchange-correlation
approximation. Hence, in the present work we focus only on the
close-packed surfaces of $4d$ transition metals. The bcc (011), fcc
(111) and hcp (0001) surfaces were modeled using slabs consisting of
$8$ atomic layers parallel to the surface plane. The slabs were
separated by vacuum layers having width equivalent with $4$ atomic
layers. The irreducible part of the two-dimensional bcc (011)
surface Brillouin zone was sampled by $120$ $\vec k$-points, whereas
for both fcc (111) and hcp (0001) surfaces we used $240$ $\vec
k$-points. The surface energy was calculated from the slab energy
and the corresponding bulk energy as described, e.g., in Ref.
\cite{Kollar2003}.

\begin{table}
\caption{Comparison between the errors in the equilibrium lattice
constants for a few selected metals calculated using the present
approach (EMTO) and those reported in Ref.\cite{Perdew2007} (in
parentheses). The mean errors (upper panel) and mean absolute errors
(lower panel) are shown for LDA, PBE and PBEsol functionals (in
units of Bohr$\times 10^{-2}$).}
\begin{ruledtabular}
\begin{tabular}{lccc}
 & LDA & PBE &PBEsol\\
 & \multicolumn{3}{c}{mean error}\\
\hline
Li, Na, K, Al & -21.4  &  3.6  &  -2.1  \\
               &(-17.0) & (5.5) & (-0.6) \\
Cu, Rh, Pd, Ag & -7.8  &  12.9  & -0.2  \\
                    &(-7.6) & (12.1) & (0.0) \\
\hline
 & \multicolumn{3}{c}{mean absolute error}\\
Li, Na, K, Al &  21.4  &  5.7  &  2.6  \\
               & (17.0) & (6.4) & (4.3) \\
Cu, Rh, Pd, Ag &  7.8  &  12.9  &  2.1 \\
                    & (7.6) & (12.1) & (3.6) \\
\end{tabular}
\end{ruledtabular}
\label{table1}
\end{table}

\begin{table}
\caption{Theoretical (EMTO) and experimental \cite{Young1991}
equilibrium atomic radii ($w$ in Bohr) and bulk moduli ($B$ in GPa)
for cubic $s$ and $p$ metals. The theoretical values are shown for
the LDA, PBE, PBEsol, LAG and AM05 functionals. The unit for the
mean absolute error ($mae$) is Bohr$\times 10^{-2}$ for $w$ and GPa
for $B$. For each element, the best results are shown in boldface.}
\begin{ruledtabular}
\begin{tabular}{cccccccc}
     &      & LDA & PBE & PBEsol & LAG & AM05 & Expt. \\
\hline
Li  & $w$  & 3.13 & 3.20 & 3.20 & 3.21 & \textbf{3.22}  & 3.237\\
fcc & $B$  & 14.0 & 13.7 & 13.8 & \textbf{13.5} & \textbf{13.5}  & 12.6 \\
\hline
Na  & $w$  & 3.77 & 3.91 & 3.89 & \textbf{3.92} & \textbf{3.92} & 3.928 \\
bcc & $B$  & 8.56 & 7.88 & 7.82 & \textbf{7.54} & 7.73 & 7.34 \\
\hline
K   & $w$  & 4.69 & 4.92 & \textbf{4.86} & 4.92 & 4.92 & 4.871\\
bcc & $B$  & 3.94 & 4.06 & 3.94 & \textbf{3.88} & 3.97 & 3.70 \\
\hline
Rb  & $w$  & 5.00 & 5.27 & \textbf{5.18} & 5.26 & 5.27 & 5.200 \\
bcc & $B$  & 3.21 & 3.34 & 3.22 & \textbf{3.19} & 3.28 & 2.92 \\
\hline
Cs  & $w$  & 5.36 & 5.73 & \textbf{5.60} & 5.72 & 5.74 & 5.622 \\
bcc & $B$  & \textbf{2.08} & 2.32 & 2.15 & 2.16 & 2.24 & 2.10 \\
\hline
Ca  & $w$  & 3.95 & \textbf{4.09} & 4.04 & 4.06 & 4.07 & 4.109 \\
fcc & $B$  & \textbf{17.9} & 16.8 & 17.1 & 16.5 & 17.0 & 18.4 \\
\hline
Sr  & $w$  & 4.30 & \textbf{4.45} & 4.38 & 4.41 & 4.42 & 4.470\\
fcc & $B$  & 14.0 & 13.2 & 13.5 & \textbf{13.1} & 13.3 & 12.4 \\
\hline
Ba  & $w$  & 4.38 & \textbf{4.67} & 4.52 & 4.59 & 4.61 & 4.659\\
bcc & $B$  & \textbf{8.29} & 7.76 & 7.72 & 7.57 & 7.49 & 9.30 \\
\hline
Al  & $w$  & 2.95 & \textbf{2.99} & 2.97 & 2.98 & 2.96 & 2.991\\
fcc & $B$ & 81.2 & \textbf{75.7} & 80.1 & 76.5 &  84.8 & 72.8 \\
\hline
Pb  & $w$  & 3.60 & 3.71 & 3.64 & \textbf{3.67} & 3.64 & 3.656\\
bcc & $B$  & 59.4 & \textbf{41.2} & 53.0 & 46.5 & 50.1 & 41.7\\
\hline \hline
$mae$ & $w$  & 16.13 & \textbf{3.87} & 4.63 & 4.45 & 4.47 &  \\
      & $B$  & 3.24 & \textbf{1.01} & 2.49 & 1.44 & 2.66 &  \\
\end{tabular}
\end{ruledtabular}
\label{table2}
\end{table}

\begin{table}[b]
\caption{Theoretical (EMTO) and experimental \cite{Young1991}
equilibrium atomic radii ($w$ in Bohr) and bulk moduli ($B$ in GPa)
for cubic $3d$ metals. For notations see caption for Table
\ref{table2}.}
\begin{ruledtabular}
\begin{tabular}{cccccccc}
     &      & LDA & PBE & PBEsol & LAG & AM05 & Expt. \\
\hline
V   & $w$  & 2.72 & \textbf{2.79} & 2.75 & 2.76 & 2.75  & 2.813 \\
bcc & $B$ & 199 & \textbf{176}  & 188 & 183    & 187 & 155 \\
\hline
Cr  & $w$  & 2.60 & \textbf{2.65} & 2.62 & 2.62 & 2.62 & 2.684\\
bcc & $B$ & 285 & \textbf{259} & 274 & 268  & 273 & 160 \\
\hline
Fe  & $w$  & 2.56 & \textbf{2.64} & 2.60 & 2.60 & 2.60 & 2.667\\
bcc & $B$ & 245 & \textbf{191} & 220 & 213  & 223 & 163 \\
\hline
Ni  & $w$  & 2.53 & \textbf{2.61} & 2.56 & 2.57 & 2.56 & 2.602\\
bcc & $B$ & 243 & \textbf{198} & 223 & 214  & 222 & 179 \\
\hline
Cu  & $w$  & 2.60 & 2.69 & 2.64 & \textbf{2.65}  & 2.64 & 2.669\\
fcc & $B$ & 182 & \textbf{142}  & 165 & 155 & 163 & 133 \\
\hline \hline
$mae$ & $w$  & 8.50 & \textbf{2.26} & 5.30  & 4.70 & 5.30 & \\
  & $B$ & 72.80  & \textbf{35.20}  & 56.00  & 48.60 & 55.6 & \\
\end{tabular}
\end{ruledtabular}
\label{table3}
\end{table}

First, we address the accuracy of the present total energy method by
comparing the EMTO results for the equilibrium lattice constant of a
few selected metals with those reported in Ref.\cite{Perdew2007}.
The latter results were generated by the Gaussian code (GC)
\cite{Staroverov2004}. The errors from Table \ref{table1} represent
the differences between the theoretical results and the experimental
data corrected for the zero-point expansion \cite{Staroverov2004}.
We find that, on average, the errors obtained using the two methods
are close to each other and follow the same trend when going from
LDA to gradient corrected approximations. The deviation between the
EMTO and the GC errors is somewhat larger for the simple metals,
which may be attributed to the fact that these solids have very
shallow energy minimum (small bulk modulus) and thus require a
higher accuracy for the EOS fitting. The overall good agreement
between the two sets of errors qualify for using the EMTO approach
to shed light on the performance of the PBEsol functional in the
case of metallic systems.

\begin{table}[t]
\caption{Theoretical (EMTO) and experimental \cite{Young1991}
equilibrium atomic radii ($w$ in Bohr) and bulk moduli ($B$ in GPa)
for $4d$ metals. For notations see caption for Table \ref{table2}.}
\begin{ruledtabular}
\begin{tabular}{cccccccc}
     &      & LDA & PBE & PBEsol & LAG & AM05 & Expt. \\
\hline
Y   & $w$  & 3.65 & \textbf{3.77} & 3.71 & 3.72 & 3.72 & 3.760\\
hcp & $B$ & \textbf{40.7} & 36.5 & 38.2 & 37.1 & 37.5 &  41.0 \\
\hline
Zr  & $w$  & 3.28 & \textbf{3.36} & 3.31 & 3.32 & 3.32 & 3.347\\
hcp & $B$ & 98.5 & 89.9 & 93.0 & 92.2 & \textbf{93.1} & 94.9\\
\hline
Nb  & $w$  & 3.01 & \textbf{3.08} & 3.04 & 3.05 & 3.04 & 3.071\\
bcc & $B$ & 171 & 146 & 160 & 154  & \textbf{162} & 169 \\
\hline
Mo  & $w$  & 2.90 & 2.94 & 2.91 & \textbf{2.92} & 2.91 & 2.928\\
bcc & $B$ & 272 & 247 & \textbf{263} & 256  & \textbf{263} & 261 \\
\hline
Tc  & $w$  & 2.82 & 2.86 & 2.83 & \textbf{2.84} & 2.83 & 2.847\\
hcp & $B$ & 323 & 286 & 310 & \textbf{301}  & 312 & 297 \\
\hline
Ru  & $w$  & 2.77 & 2.82 & 2.79 & \textbf{2.80} & 2.78 & 2.796\\
hcp & $B$ & 353 & \textbf{305} & 336 & 325 & 339 & 303 \\
\hline
Rh  & $w$  & 2.78 & 2.84 & \textbf{2.80} & 2.81 & \textbf{2.80} & 2.803\\
fcc & $B$ & 304 & 251  & \textbf{285} & 272  & 286 & 282 \\
\hline
Pd  & $w$  & \textbf{2.85} & 2.92 & 2.87 & 2.89 & 2.87 & 2.840\\
fcc & $B$ & 229 & 166  & 205 & \textbf{191} & 204 & 189 \\
\hline
Ag  & $w$  & 2.97 & 3.07 & 3.00 & 3.03 & \textbf{3.01} & 3.018 \\
fcc & $B$ & 137 & 89.6 & 117 & \textbf{106}  & 110 & 98.8\\
\hline \hline
$mae$  & $w$  & 4.44 & 2.78 & 2.33 & \textbf{1.96} & 2.11 &  \\
   & $B$ & 21.46 & 13.63  & 10.99  & \textbf{7.98} & 10.6 &  \\
\end{tabular}
\end{ruledtabular}
\label{table4}
\end{table}

\begin{table}[b]
\caption{Theoretical (EMTO) and experimental \cite{Young1991}
equilibrium atomic radii ($w$ in Bohr) and bulk moduli ($B$ in GPa)
for cubic $5d$ metals. For notations see caption for Table
\ref{table2}. }
\begin{ruledtabular}
\begin{tabular}{cccccccc}
     &      & LDA & PBE & PBEsol & LAG & AM05 & Expt. \\
\hline
Ta  & $w$  & 3.03 & 3.10 & 3.06 & \textbf{3.07} & 3.06 & 3.073 \\
bcc & $B$ & \textbf{194} & 180 & 188 & 183  & \textbf{188} & 191 \\
\hline
W   & $w$  & 2.92 & 2.97 & \textbf{2.94} & 2.95 & \textbf{2.94} & 2.937\\
bcc & $B$ & \textbf{308} & 294 & 305 & 298  & 307 & 308 \\
\hline
Ir  & $w$  & 2.83 & 2.87 & \textbf{2.84} & 2.85 & \textbf{2.84} & 2.835\\
fcc & $B$ & 392 & 340 & 376 & \textbf{362}  & 382 & 358 \\
\hline
Pt  & $w$  & 2.89 & 2.95 & 2.91 & 2.92 & \textbf{2.90} & 2.897 \\
fcc & $B$ & 299 & 243 & \textbf{281} & 265 & 283  & 277 \\
\hline
Au  & $w$  & \textbf{3.00} & 3.08 & 3.03 & 3.05 & 3.02 & 3.013\\
fcc & $B$ & 188 & 136 & 170 & 155  & \textbf{168} & 166  \\
\hline \hline
$mae$ & $w$  & 1.70 & 4.30 & 1.02 & 1.82 & \textbf{0.62} &   \\
  & $B$ & 16.20  & 21.40  & \textbf{6.40}  & 9.00 & 7.20 & \\
\end{tabular}
\end{ruledtabular}
\label{table5}
\end{table}

Next, we discuss the present results obtained for the equation of
state. In Table \ref{table2}, we list the EMTO equilibrium atomic
radii ($w$) and bulk moduli ($B$) for monovalent $sp$ metals (Li,
Na, K, Rb, Cs), cubic divalent $sp$ metals (Ca, Sr, Ba) and for Al
and Pb. Tables \ref{table3} and \ref{table5} show results for the
cubic $3d$ and $5d$ metals, respectively, whereas in Table
\ref{table4} we give results for the entire $4d$ series. The mean
absolute errors ($mae$) for $w$ and $B$ calculated in LDA, PBE,
PBEsol, LAG and AM05 approximations are shown at the bottom of the
tables.

As expected, for all metals the LDA underestimation of the
equilibrium volume is reduced by the gradient corrected functionals.
This is especially pronounced in the case of simple metals and $3d$
transition metals. When comparing the performances of the four
gradient level approximations, we find similar errors for the simple
metals and $4d$ transition metals. PBE yields far the best volumes
for the $3d$ metals, whereas the volumes of the $5d$ metals are best
described by AM05 followed by PBEsol. For simple metals, we have the
following sequence: $w({\rm LDA})< w({\rm PBEsol})< w({\rm
LAG})\lesssim w({\rm AM05})$ (except Al and Pb) and $w({\rm
AM05})\lesssim w({\rm PBE})$ (except Li). For all transition metals
and also for Al and Pb, we have: $w({\rm LDA})< w({\rm
PBEsol})\approx w({\rm AM05})\lesssim w({\rm LAG})<w({\rm PBE})$.
Surprisingly, for most of the metals the PBEsol atomic radii are
only slightly smaller than those obtained within the LAG
approximation: the average difference being $\sim 0.0018$ Bohr for
the simple metals and $\sim 0.0012$ Bohr for the transition metals.
For comparison, the corresponding differences between the PBEsol and
PBE radii are $\sim 0.0076$ Bohr and $\sim 0.0028$ Bohr. This
finding indicates that the surface-like correlation effects (present
in PBEsol and AM05 but neglected in LAG) play minor role in the bulk
equilibrium properties of metals.

\begin{table}
\caption{Theoretical surface energies (in J/m$^{2}$) for the
close-packed surfaces of $4d$ transition metals. Results are shown
for the LDA, PBE, PBEsol, LAG and AM05 functionals. For comparison,
the results for Rb and Sr are also included.}
\begin{ruledtabular}
\begin{tabular}{llccccc}
  & surface & LDA & PBE & PBEsol & LAG & AM05\\
\hline
Rb & bcc (110)  & 0.12 & 0.09 & 0.11 & 0.08 & 0.09 \\
Sr & fcc (111)  & 0.55 & 0.44 & 0.50 & 0.44 & 0.47 \\
Y  & hcp (0001) & 1.38 & 1.18 & 1.31 & 1.18 & 1.30 \\
Zr & hcp (0001) & 2.15 & 1.90 & 2.08 & 1.89 & 2.04 \\
Nb & bcc (110)  & 2.66 & 2.32 & 2.58 & 2.30 & 2.55 \\
Mo & bcc (110)  & 3.69 & 3.23 & 3.59 & 3.24 & 3.58 \\
Tc & hcp (0001) & 3.86 & 3.25 & 3.70 & 3.35 & 3.74 \\
Ru & hcp (0001) & 4.18 & 3.47 & 3.99 & 3.62 & 3.98 \\
Rh & fcc (111)  & 3.34 & 2.63 & 3.14 & 2.80 & 3.11 \\
Pd & fcc (111)  & 2.29 & 1.65 & 2.08 & 1.80 & 2.02 \\
Ag & fcc (111)  & 1.40 & 0.89 & 1.23 & 1.13 & 1.13 \\
\end{tabular}
\end{ruledtabular}
\label{table6}
\end{table}

The sensitivity of the bulk modulus to the exchange-correlation
approximation is similar to that of the atomic radius. PBE gives the
smallest $mae(B)$ for the simple and $3d$ metals, while the $4d$ and
$5d$ metals have the lowest $mae(B)$ for LAG and PBEsol,
respectively. Except a few simple metals, we find $B({\rm
LDA})>B({\rm PBEsol})\approx B({\rm AM05})>B({\rm LAG})>B({\rm
PBE})$. The large PBE errors in $B$ for the late $5d$ metals are
greatly reduced by the PBEsol and AM05 approximations.
Unfortunately, both the atomic radii and bulk moduli of magnetic
$3d$ metals are very poorly described by the present approximations.

In order to be able to judge the relative merits of the four
gradient level functionals for bulk systems, we consider the mean
absolute errors for all 29 metals from Tables
\ref{table2}-\ref{table5}. We find that the total $mae$'s for $w$
are $8.69$, $3.32$, $3.41$, $3.27$ and $3.22$ Bohr$\times10^{-2}$ in
LDA, PBE, PBEsol, LAG and AM05, respectively. The same figures for
$B$ are $23.12$, $14.34$, $15.03$, $12.90$ and $15.03$ GPa. Thus,
AM05 yields marginally better $w$ and LAG marginally better $B$
compared to the other gradient approximations. However, this
comparison is meaningful only within the error bar associated with
the particular computational method. Using the GC and EMTO
\emph{mae}'s from Table \ref{table1} and assuming a hypothetical fcc
structure for all metals from this table, for the average $mae$ in
$w$ we obtain $3.5$ Bohr$\times10^{-2}$ for GC and $3.4$
Bohr$\times10^{-2}$ for EMTO. The deviation between the two average
$mae$'s sets the error of the EMTO equilibrium radii to $\pm 0.1$
Bohr$\times10^{-2}$. For the error of the EMTO bulk moduli we use
$\pm 2$ GPa, which is the error associated with the present
perturbative treatment of the gradient terms. Taking into account
these error bars, we conclude that for bulk metals the PBEsol
functional has the accuracy of the PBE, LAG and AM05 functionals.

\begin{figure}[t]
\includegraphics[height=8cm]{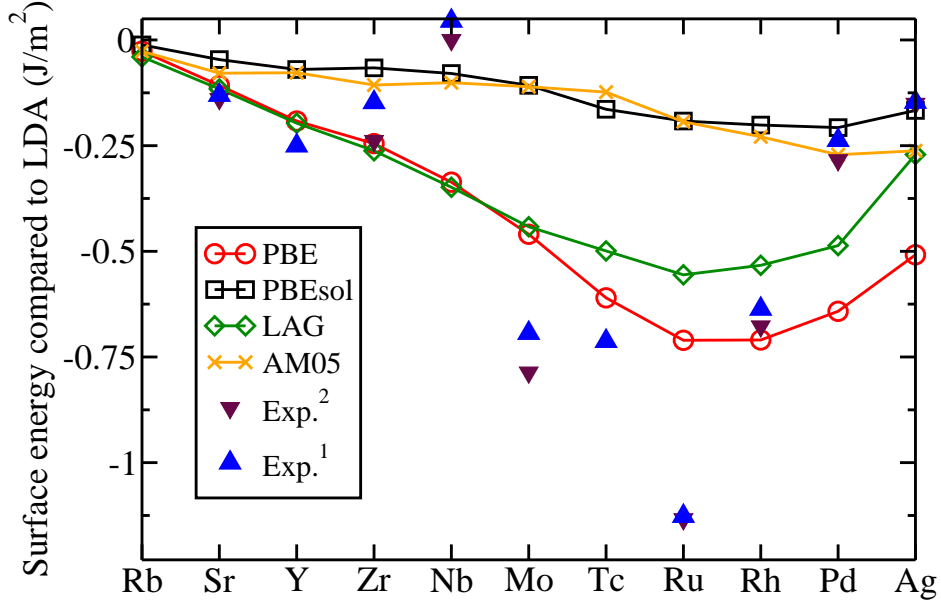}
\caption{(Color online) The effect of PBE (red circles), PBEsol
(black squares), LAG (green diamonds) and AM05 (yellow crosses)
gradient corrections on the LDA surface energies for Rb, Sr and $4d$
transition metals (in J/m$^2$). For comparison, the differences
between the experimental surface energies (blue triangle up:
Expt.$^1$ Ref. \cite{Tyson1977}; maroon triangle down: Expt.$^2$
\cite{Boer1988}) and LDA values are also shown.} \label{fig1}
\end{figure}

In the following, we discuss the surface energy ($\gamma$)
calculated for the close-packed surfaces of $4d$ transition metals.
The EMTO surface energies $\gamma_{xc}$ ($xc$ stands for LDA, PBE,
PBEsol, LAG or AM05) are listed in Table \ref{table6}. To illustrate
the effect of different gradient corrections, in Figure \ref{fig1}
we show the surface energy differences $\Delta \gamma_{xc}\equiv
(\gamma_{xc}-\gamma_{\rm LDA})$. For completeness, the differences
between the experimental \cite{Tyson1977,Boer1988} and LDA surface
energies have also been included in figure.

Today, the most comprehensive experimental surface energy data is
the one derived from the surface tension measurement in the liquid
phase and extrapolated to zero temperature
\cite{Tyson1977,Boer1988}. Using these experimental data, for the
mean absolute values of the relative errors we get $18.7\%$ for LDA,
$11.2\%$ for PBE, $12.9\%$ for PBEsol, $10.7\%$ for LAG, and
$14.1\%$ for AM05. This would place the PBE approximation on the top
followed by the LAG, PBEsol, AM05 and LDA. However, the accuracy of
the experimental surface energies at low temperature is not known
and therefor a direct comparison of the absolute values of
$\gamma_{xc}$ to the experimental data is not suitable for
establishing the performance of different functionals for metal
surfaces. Because of that, in the following we investigate the
effect of gradient corrections relative to LDA.

From Table \ref{table6} and Figure \ref{fig1}, we see that the
gradient correction always decreases the surface energy. Except Rb,
the theoretical surface energies follow the trend $\gamma_{\rm LDA}>
\gamma_{\rm PBEsol}\approx \gamma_{\rm AM05} >\gamma_{\rm
LAG}\gtrsim\gamma_{\rm PBE}$. It is clear that PBE has large
negative effect on the surface energies: its relative effect
$\delta_{\rm PBE}\equiv |\Delta \gamma_{\rm PBE}|/\gamma_{\rm PBE}$
is ranging between $\sim 12\%$ (Zr and Mo) and $\sim 38\%$ (Ag). The
effect of LAG is somewhat smaller in late $4d$ metals compared to
that of PBE. The situation is very different for the PBEsol and AM05
functionals. First, these two approximations lead to a rather
uniform change relative to the LDA surface energies. Second,
$\delta_{\rm PBEsol}$ and $\delta_{\rm AM05}$ remain below $\sim
8\%$ for most metals, except Rb, Pd and Ag, where the PBEsol (AM05)
gradient effect reaches $\sim 12\%$ ($\sim 20\%$) of the LDA surface
energy. We point out the the large $(\gamma_{\rm AM05}-\gamma_{\rm
LAG})$ values calculated for Nb, Mo, Tc, Ru and Rh are due to the
surface-like correlation effects neglected in the LAG approach.
While such effects are small for bulk simple metals (Table
\ref{table2}) and almost negligible in bulk transition metals
(Tables \ref{table3}-\ref{table5}), they can be as large as $\sim
0.4$ J/m$^2$ (or $\sim 10\%$ of the LDA surface energy), obtained
for the hcp (0001) surface of Tc.

We recall that the surface energy of jellium surfaces has been found
to be more accurately described in LDA than in GGA
\cite{Perdew1992a}. Furthermore, it has recently been shown that LDA
yields surface energies of ceramics in better agreement with the
broken bond model than GGA \cite{Liu2006}. This is surprising,
especially taking into account that the broken bond model is based
on the cohesive energy, which can be calculated accurately within
GGA. On this ground, one tends to assume that the LDA surface
energies are closer to the true surface energies than the PBE ones.
Considering the relatively small effect of the PBEsol and AM05
approximations over the LDA surface energies (Figure \ref{fig1}), it
is likely that these two functionals perform better for metallic
surfaces compared to PBE and LAG.

We conclude that for metallic bulk and surface systems, the newly
developed PBEsol approximation, belonging to the generalized
gradient approximation (GGA) family of exchange-correlation
functionals, has the accuracy of the AM05 functional derived from
model subsystem (SSF) approach. Based on the assumption that the
true surface energy of transition metals is close to the LDA surface
energy, we suggest that these two functionals are superior compared
to former gradient level approximations.

\emph{Acknowledgments:} The Swedish Research Council, the Swedish
Foundation for Strategic Research, the Academy of Finland (No.
116317) and the Hungarian Scientific Research Fund (T046773 and
T048827) are acknowledged for financial support. M.R. and K. K.
acknowledges the computer resources of the Finnish IT Center for
Science (CSC) and Mgrid project.

\end{document}